\documentclass[english,reprint,twocolumn,amsmath,amssymb,prl,nofootinbib,superscriptaddress]{revtex4-1}
\usepackage[T1]{fontenc}
\usepackage[latin9]{inputenc}
\usepackage{geometry}
\usepackage{graphicx}
\usepackage{bm}
\usepackage{hyperref}
\hypersetup{colorlinks, citecolor=Blue, linkcolor=Blue, urlcolor=Blue}
\usepackage[normalem]{ulem}
\usepackage[usenames,dvipsnames]{color}
\usepackage[sort&compress]{natbib}
\usepackage{textcomp}
\usepackage{babel}
\usepackage{notes2bib}
\bibnotesetup{
note-name = ,
use-sort-key = false
}

\AtBeginDocument{
   \setlength\abovedisplayskip{6pt}
   \setlength\belowdisplayskip{6pt}}

\begin{document}
\setlength{\parskip}{0pt}
\title{Probing Dark Matter Using Precision Measurements of Stellar Accelerations}

\author{Aakash~Ravi}
\affiliation{Department of Physics, Harvard University, Cambridge, MA 02138, USA}
\affiliation{Harvard-Smithsonian Center for Astrophysics, 60 Garden St., Cambridge, MA 02138, USA}

\author{Nicholas~Langellier}
\affiliation{Department of Physics, Harvard University, Cambridge, MA 02138, USA}
\affiliation{Harvard-Smithsonian Center for Astrophysics, 60 Garden St., Cambridge, MA 02138, USA}

\author{David~F.~Phillips}
\affiliation{Harvard-Smithsonian Center for Astrophysics, 60 Garden St., Cambridge, MA 02138, USA}

\author{Malte~Buschmann}
\affiliation{Leinweber Center for Theoretical Physics, Department of Physics, University of Michigan, Ann Arbor, Michigan 48109, USA}

\author{Benjamin~R.~Safdi}
\affiliation{Leinweber Center for Theoretical Physics, Department of Physics, University of Michigan, Ann Arbor, Michigan 48109, USA}

\author{Ronald~L.~Walsworth}
\email{rwalsworth@cfa.harvard.edu}
\affiliation{Department of Physics, Harvard University, Cambridge, MA 02138, USA}
\affiliation{Harvard-Smithsonian Center for Astrophysics, 60 Garden St., Cambridge, MA 02138, USA}

\date{\today}
\begin{abstract}
Dark matter comprises the bulk of the matter in the universe but its particle nature and cosmological origin remain mysterious. Knowledge of the dark matter density distribution in the Milky Way Galaxy is crucial to both our understanding of the standard cosmological model and for grounding direct and indirect searches for the particles comprising dark matter. Current measurements of Galactic dark matter content rely on model assumptions to infer the forces acting upon stars from the distribution of observed velocities. Here, we propose to apply the precision radial velocity method, optimized in recent years for exoplanet astronomy, to measure the change in the velocity of stars over time, thereby providing a direct probe of the local gravitational potential in the Galaxy. Using numerical simulations, we develop a realistic strategy to observe the differential accelerations of stars in our Galactic neighborhood with next-generation telescopes, at the level of $10^{-8}$ cm/s$^{2}$. Our simulations show that detecting accelerations at this level with an ensemble of 10$^3$ stars requires the effect of stellar noise on radial velocity measurements to be reduced to <10 cm/s. The measured stellar accelerations may then be used to extract the local dark matter density and morphological parameters of the density profile.
\end{abstract}
\maketitle

\section{Introduction}
Understanding the nature of dark matter (DM)~\cite{Bertone2016} is one of the most pressing issues in modern physics. Many particle DM models, such as those employing weakly-interacting massive particles and axions, predict unique laboratory and astrophysical signatures, which are being searched for at a variety of experiments and observatories \cite{Patrignani:2016xqp}.  However, knowledge of the local DM density is crucial for interpreting the results of these efforts. Unfortunately, current methods for determining the local properties of DM ({\it i.e.}, within our region of the Galaxy) are indirect and subject to large systematic uncertainties~\cite{Read2014}.  In addition to aiding searches for particle DM, better certainty of the local DM distribution may provide key insights into the history of the Milky Way (MW).  In this work, we propose a new approach -- direct measurements of stellar accelerations -- to determine the local DM density and morphological parameters in the MW.  Our technique circumvents many of the systematic issues faced by existing methods.

Currently, the DM density in the MW is inferred from either the Galactic rotation curve, measured via Doppler shifts, or the dispersion of local stellar velocities in the vertical direction about the Galactic mid-plane~\cite{Kafle2012,Bovy:2012tw,Read2014,Pato:2015dua,Schutz2018}, measured using astrometry. However, implicit in both these analyses is the assumption of equilibrium, {\it i.e.}, that dynamics have reached steady-state. In particular, the velocity distribution does not directly probe the gravitational potential and thus the DM distribution: only the equilibrium velocity distribution is determined by the potential. Given the presence of density waves in the MW, for example those causing the local North-South asymmetry recently studied in~\cite{2019MNRAS.482.1417B} using \emph{Gaia}, as well as other out-of-equilibrium processes, such as the continuing interactions of the Galaxy with massive satellites, the equilibrium assumption used in traditional determinations of the local DM density is open to question. Stellar accelerations, on the other hand, are directly determined by the forces acting on a star, and thus the gravitational potential, with no modeling assumptions.  Having such a direct probe would allow, in principle, for an unbiased mapping of the gravitational potential of the MW.  This approach opens up, for example, the possibility of searching for low-mass DM subhalos, which are predicted in the standard cosmological framework but are absent in certain DM models such as warm DM and fuzzy DM \cite{Buckley:2017ijx}.
 
Stellar accelerations may also be used to probe both the spatial morphology of the density profile of the bulk DM halo and the Galactic disk, in addition to possible DM subhalos (see~\cite{Siegel:2007fz,2013ApJ...767....9H,Hezaveh:2014aoa,Erkal:2014tda,2016MNRAS.463..102E,VanTilburg:2018ykj} for related but more indirect proposals). The radial density profile in particular plays a key role in interpreting searches for DM annihilation \cite{Bertone2005}, while the halo shape is influenced by baryonic feedback and potentially DM self-interactions~\cite{Tulin2018}.  

In this Letter, we show how precision radial velocity (RV) measurements of the accelerations of individual stars in our Galactic neighborhood can be used to measure the local DM density and constrain the spatial morphology of the DM density distribution. To our knowledge, a similar idea has only been suggested once before, in the context of testing modified Newtonian gravity with MW globular clusters \cite{Amendola2008,Quercellini2008}. We highlight the work of Silverwood and Easther \cite{Silverwood2018}, contemporaneous to our own, in which they also propose using stellar accelerations to map the Galactic gravitational field, and provide a complementary analysis. 

\begin{figure} 
\includegraphics[width=\columnwidth]{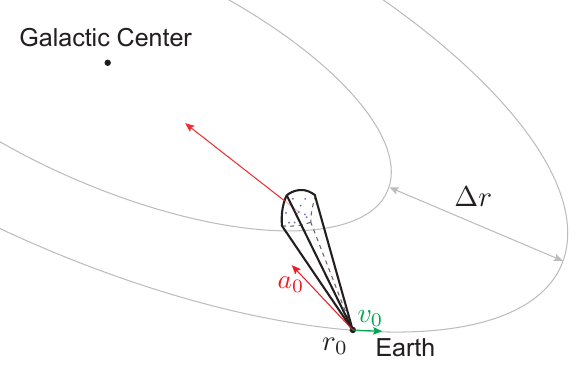} 
\caption{\label{fig:schematic}Geometry for observing stellar accelerations in the Milky Way. The solar system is at a distance $r_0$ from the Galactic Center (origin), has a rotational velocity $v_0$ and feels an acceleration $a_0$ due to the Milky Way gravitational potential. Stars further inward feel a stronger acceleration. From Earth, we can observe the radial velocity of stars $\Delta r$ away. By measuring small changes in these velocities over time, we directly determine stellar accelerations and hence the Milky Way gravitational potential. The diagram above is not to scale and angles are exaggerated for effect.} 
\end{figure}

We propose to use the precision RV method, honed in the search for extrasolar planets \cite{Fischer2016}, to measure directly the change over time of the velocities of an ensemble of individual stars -- and thus the forces acting upon those stars. Since the Sun is also accelerating in the gravitational potential of the MW, measurements must be performed on stars sufficiently distant from the Sun -- either closer or further from the Galactic Center (GC) -- for a difference in acceleration to be observed. Given the rotational velocity of the Sun about the GC $v_{\text{circ}}\left(r_0\right) \equiv v_0 \approx 220$~km/s \cite{Bovy2012}, and our Galactocentric distance $r_0 \approx 8$~kpc \cite{Bovy2012}, the local centripetal acceleration is $a_r\left(r_0\right)\equiv a_0\approx 2\times 10^{-8}$~cm/s$^2$. This situation is depicted in Figure \ref{fig:schematic}.

\section{Theoretical framework}

The Poisson equation directly relates stellar acceleration gradients, which we propose to measure using the RV method, to the energy density $\rho$: $\nabla \cdot {\bf a} = -4 \pi G \rho$, where $G$ is Newton's constant. Suppose we measure the radial acceleration gradients for stars towards the GC, as pictured in Fig.~\ref{fig:schematic}. Note that our proposed observing region will be intentionally slightly misaligned from the GC (vertically and horizontally) to avoid extinction in the Galactic midplane and overcrowding of stars.  To a good approximation, pointing in this manner primarily gives us a measure of $\partial a_r/\partial r$, where $a_r$ is the component of the acceleration in the radial direction. We assume azimuthal symmetry and discuss errors from contamination by vertical gradients later in this section.

Radial gradients of the acceleration are primarily determined by the local DM density. Using the Poisson equation, we may relate the radial gradients of $a_r$ to the local DM density, finding
\begin{align}
\label{eq: key}
\rho_{\text{DM}} \approx \frac{1}{4\pi G}\left(2(A-B)^{2} - \frac{\partial a_{r}}{\partial r}\right) \,,
\end{align}
where $A = 15.3 \pm 0.4$ km s$^{-1}$ kpc$^{-1}$ and $B = -11.9 \pm 0.4$  km s$^{-1}$ kpc$^{-1}$ are the Oort constants~\cite{2017MNRAS.468L..63B}.
Note that the combination of Oort constants in Eq.~\eqref{eq: key} is related to the circular velocity and distance from the GC by $A - B = v_0 / r_0$. The relation in Eq.~\eqref{eq: key} would be exactly true were it not for the contribution of the Galactic disk, which contributes energy density locally.  However, even though the local energy density due to the disk is expected to dominate that of the DM by approximately a factor of 10~\cite{Read2014}, we find that completely neglecting the disk leads to only a $30$\% error in measuring the DM density using~\eqref{eq: key}.  At higher precision, the contribution from the disk can be modeled, as we discuss below.

First, it is instructive to understand Eq.~\eqref{eq: key} in the context of a spherical DM density profile.  In this case, we may write the DM contribution to the acceleration as $a_r(r) = -G \, M(r) / r^2$, where $M(r)$ is the DM mass enclosed within the radius $r$.  Note that the contribution to the Oort constants from the spherical potential is $2(A-B)^2 = 2 G\, M(r) /r^3$, while the derivative gives $a'(r) = 2 G \, M(r) / r^3 - G \, M'(r) / r^2$.  Defining the local DM density as $\rho_{\rm DM}$ and $r_0$ as the distance from the Sun to the GC, we then find that the right hand side of Eq.~\eqref{eq: key} trivially reduces to $\rho_{\rm DM}$ since $M'(r_0) =4 \pi r_0^2 \rho_{\rm DM}$.   

Now let us repeat the exercise above for the disk density profile in order to calculate the contribution to Eq.~\eqref{eq: key} from the disk, which should in principle be subtracted at high enough precision.  For a thin disk with surface density $\Sigma_{\rm disk}$, and assuming that we are in the plane of the disk, the calculation proceeds as for the spherical density profile except that $M'_{\rm disk}(r) = 2 \pi r \Sigma_{\rm disk}$, with $M_{\rm disk}(r)$ being the mass enclosed within the radius $r$ due to the disk.  We then see that the disk leads to a fictitious contribution to $\rho_{\rm DM}$, as inferred from Eq.~\eqref{eq: key} and which we refer to as $\rho_{\rm DM}^{\rm fict.}$, given by $\rho_{\rm DM}^{\rm fict.} = \Sigma_{\rm disk} / (2 r_0)$.  The local disk surface density is measured to be $\Sigma_{\rm disk} \approx 50$ $M_\odot/$pc$^2$~\cite{Read2014,Schutz2018}, with approximately and conservatively 20\% uncertainty, and also $r_0 \approx 8$ kpc. Here $M_\odot$ is the mass of the Sun. This implies that $\rho_{\rm DM}^{\rm fict.} = (3.1  \pm 0.6) \times 10^{-3}$ $M_\odot/$pc$^3$.  Current estimates put the local DM density at $\rho_{\rm DM} \approx 0.01$ $M_\odot/{\rm pc}^3$~\cite{Read2014}, meaning that $\rho_{\rm DM}^{\rm fict.}/ \rho_{\rm DM} \approx 0.3$. 
 
The relative uncertainty in the DM density is magnified compared to the relative uncertainty in $\partial a_r /\partial r$ because of the cancellation that occurs in~\eqref{eq: key}.  Writing $\delta a_r'$ as the uncertainty in $\partial a_r /\partial r$ and $\delta \rho_{\rm DM}$ as the  uncertainty in the DM density, we find
\begin{equation}\label{eq:rhoDM_unc}
{\delta \rho_{\rm DM} \over \rho_{\rm DM}} \approx {\delta a_r' \over a_r'} { (A-B)^2 \over 2 \pi G \rho_{\rm DM} }\approx 2.7 {\delta a_r' \over a_r'} \,.
\end{equation} 

Note that while we do not explore this possibility in detail in this work, the local DM density may also be measured by using vertical acceleration measurements of stars in the local neighborhood but at sufficiently high vertical displacement from the disk $z$, such that the dominant vertical acceleration is from DM and not the disk.  Note, however, the vertical acceleration from DM is sub-dominant compared to the radial acceleration by the factor $z/r_0$.  Additionally, morphological parameters for the DM density profile, such as the local slope, may be determined from higher precision acceleration measurements. We leave projections for how well such parameters could be determined to future work.

\section{Observational considerations}

If we select stars 3 kpc away (beyond the star field observed by the \emph{Kepler} spacecraft \cite{Koch2010} but potentially observable with the Giant Magellan Telescope \cite{Baldwin2018}), the fractional change in acceleration compared to the local acceleration should be approximately 0.75, i.e., $\Delta a_r = 1.5\times 10^{-8}$~cm/s$^2$. In more convenient units, this is 0.5~cm/s/year. Over 10 years, one would then expect a typical stellar velocity change due to the MW gravitational potential of approximately 5~cm/s, which is similar to the RV amplitude associated with an Earth-like exoplanet in the habitable zone around a Sun-like star \cite{Perryman2011}. We note that next-generation instruments designed for RV studies of exoplanets, including ESPRESSO~\cite{Pepe2010} and G-CLEF~\cite{Baldwin2018}, are expecting to achieve 10 cm/s or better RV  precision and long-term stability in order to pursue such exoplanet astronomy.

At a target signal-to-noise ratio of 100, the faintness of stars sets the exposure times to approximately 12 minutes per star with a 30-m telescope at $\Delta r = 1$~kpc or with a future 100-m telescope at $\Delta r = 3$~kpc. Including the length of nights as well as star visibility leads to $\sim 10^4/N$ observations per star per year with a single telescope for a $N$-star survey where each night is time-shared between various targets \cite{SM}.

Necessary for measuring small stellar accelerations is extremely stable calibration of the spectrograph used to determine the RVs over several years. The ideal tool for this task is a laser frequency comb optimized for calibrating spectrographs. These specialized instruments, known as ``astro-combs''  \cite{Li2008,Steinmetz2008,Braje2008,Glenday2015,Probst2014,McCracken2017}, may be referenced to GPS-disciplined atomic clocks.  Thus, spectrograph wavelength solutions are easily trustworthy over a decade; and even measurements from multiple comb-calibrated observatories can be combined into a single data set if the same reference clock is used for the astro-combs at all observatories.

An additional effect which needs to be considered is the contribution to radial velocities from the motion of stars in the plane of the sky, the so-called perspective acceleration. This effect can be removed using high-precision astrometric surveys such as \emph{Gaia} \cite{Brown2018}. As pointed out by Silverwood and Easther \cite{Silverwood2018}, this effect can be subtracted out at the $\sim$ 1\% level by choosing stars below a transverse velocity threshold of $\sim$ 55 km/s, and such stars are quite abundant in the \emph{Gaia} catalogue.

\section{Simulation Scheme}

\begin{figure} 
\includegraphics[width=\columnwidth]{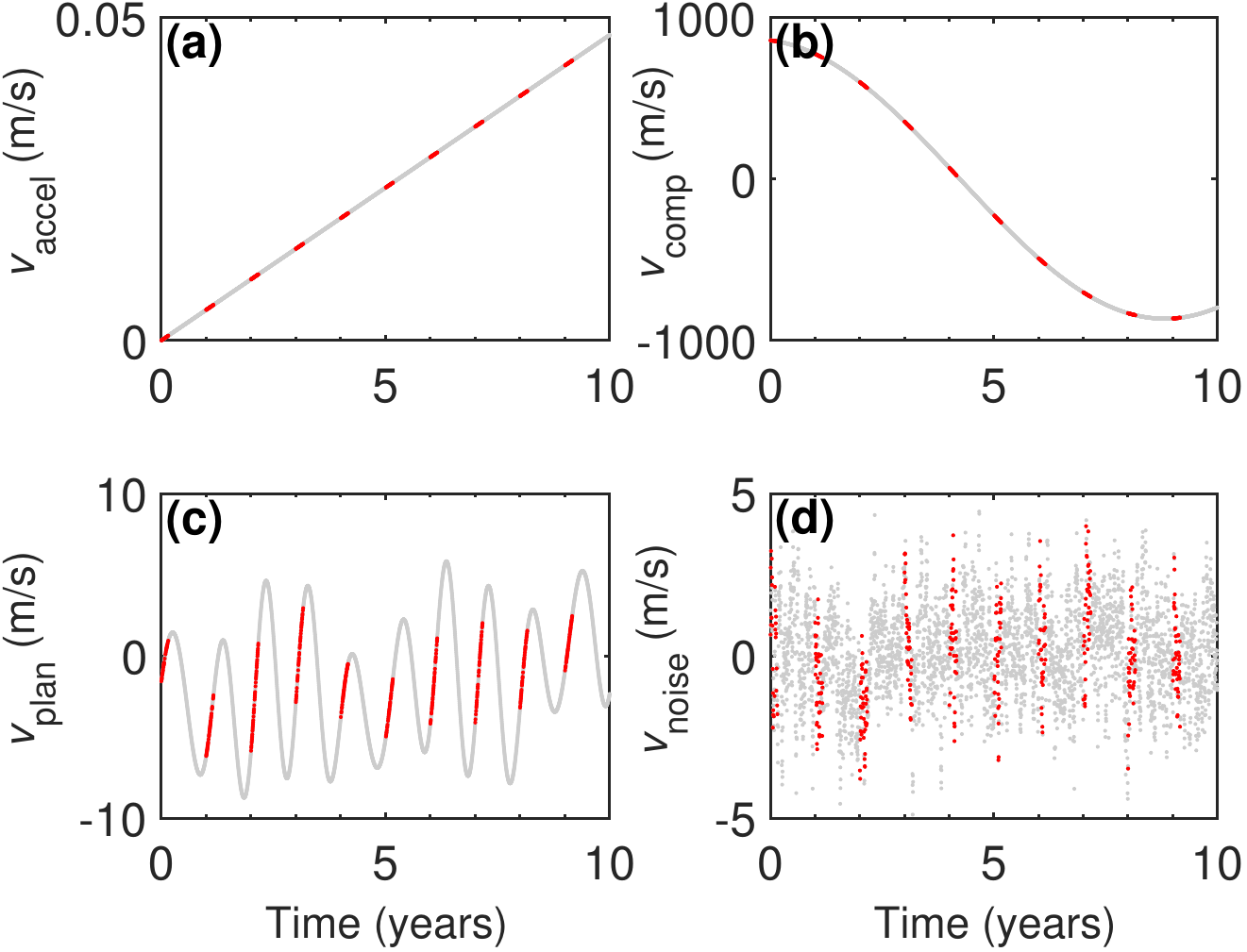} 
\caption{\label{fig:time-series}Example of a synthesized RV time series for a single primary star, showing the four contributing mechanisms considered in our analysis. The effects of (a) the acceleration due to the Milky Way gravitational potential, (b) stellar companions, (c) planets (multiple planets in this case), and (d) ``noise'' including stellar activity and instrumental effects are depicted. Red dots are the observed nights and grey dots represent all nights.}
\end{figure}

To determine whether sufficiently sensitive stellar acceleration measurements are possible, given other sources of Doppler shift ``systematics'' (e.g., stellar companions, planets, stellar noise), we simulate a measurement campaign with synthetic RV time series from a population of stars including all the above effects and a realistic observing schedule. We then analyze the time series to try to recover an injected acceleration signal of order few cm/s over a decade.

An observing schedule is generated for a total measurement campaign of 10 years \cite{SM}. The observing schedule is applied to the generation of $N$ time series ($10^3$ or larger in this Letter), representing $N$ candidate stars in an initial sample. Though the telescope time associated with the observing schedule used in this work exceeds the limits posed by the exposure times calculated in the previous section for a single telescope, we use it to aid understanding confounding effects in the search for the acceleration signal. It is important to note that fewer than $N$ stars will actually be followed in a real campaign because many stars will be poor targets for detecting an acceleration signal (due to reasons discussed below). The full campaign would then consist of a target selection phase (where one prunes the list of candidate targets) lasting a few years followed by a $\sim$decade observations devoted exclusively to measuring the acceleration signal.

Each star in the simulation is assigned a number of stellar companions and planets. The multiplicity and orbital parameters of these orbiting bodies are determined using known statistical distributions \cite{SM}. 
The total RV for a given primary star is:
\begin{equation}
v_{\textrm{total}}\left(t\right)=v_{\textrm{accel}}\left(t\right)+v_{\textrm{comp}}\left(t\right)+v_{\textrm{plan}}\left(t\right)+v_{\textrm{noise}}\left(t\right),
\end{equation}
where $v_{\textrm{accel}}\left(t\right)=\Delta a_r t$ according to the acceleration change $\Delta a_r$ at target distance $\Delta r$ away, due to the MW gravitational potential. The three terms following $v_{\textrm{accel}}$, arise from stellar companions, planets, and noise \cite{SM}. Since the survey volume is relatively small \cite{SM} and $\Delta r=3$~kpc, we take $\Delta a_r=1.5\times10^{-8}$ cm/s$^2$ ($\sim 5$ cm/s/decade) for all stars.  Figure \ref{fig:time-series} shows a typical example of each of the components plotted separately as time series; note the scale of the various effects. In this particular example, $v_{\textrm{total}}$ would be dominated by $v_{\textrm{comp}}$. The ideal case (occurring about 10\% of the time) is a star with no stellar companions or planets. This leaves us with only the MW acceleration plus noise.

\section{Results}

In a series of numerical experiments, we attempt to recover an injected $\sim 5$ cm/s/decade acceleration signal from a large ensemble of synthesized $v_{\textrm{total}}$ time series for $N$ simulated primary stars.

\begin{figure} 
\includegraphics[width=\columnwidth]{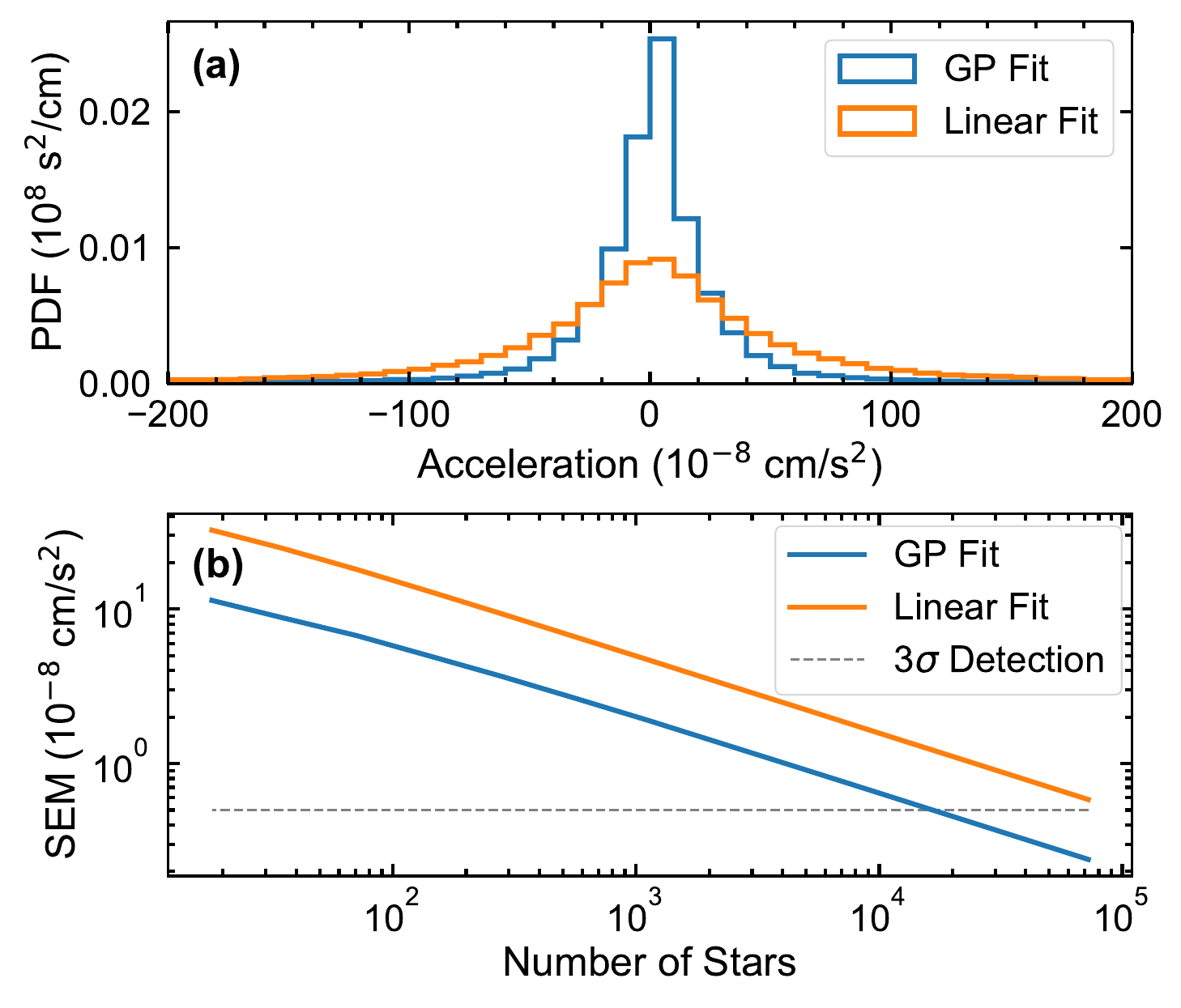} 
\caption{\label{fig:GPfit} (a) Probability density function (PDF) of fitted stellar accelerations for $\sim7.5\times10^4$ stars using Gaussian processes (GP) (blue) and a simple linear fit (orange). 60 cm/s of white noise and quasiperiodic correlated noise ranging from 50 cm/s to 250 cm/s is added to each time series. (b) Standard error of the mean (SEM) vs. number of stars observed. A 3$\sigma$ detection (dotted gray line) is obtained after about $2\times10^4$ stars for the GP fit.}
\end{figure}

First, to assess what is possible with present-day state-of-the-art instruments and analysis techniques, we set the Gaussian white noise standard deviation $\sigma_{\textrm{WN}}$ to 60~cm/s and add correlated noise with amplitudes ranging from 50 cm/s to 250 cm/s on a given star (the range of noise levels observed in the Sun \cite{Meunier2010}). In order to fit the more complicated noise model, we employ a Gaussian process (GP) regression \cite{SM} with a quasiperiodic kernel function \cite{Roberts2013,Rasmussen2006,Haywood2014}. The GP regression is also able to simultaneously fit the Keplerian components of the time series (i.e., those associated with planets and stellar companions). With the GP fit ($N = 75,\!044$ stars) we obtain a mean (standard error of the mean $\sigma$ given in parentheses) of 1.42(0.24)$\times10^{-8}$~cm/s$^2$. In contrast, a simple linear fit ($N = 72,\!425$ stars) to the time series yields a mean acceleration of 2.00(0.58)$\times10^{-8}$~cm/s$^2$. Thus the GP fit reduces the uncertainty by more than a factor of 2 compared to the linear fit, recovers the injected stellar acceleration signal of $1.5\times10^{-8}$~cm/s$^2$ within 1$\sigma$ and is $\approx6\sigma$ away from a null result. The probability density functions for the fitted accelerations are shown in Figure \ref{fig:GPfit}(a).

We can vary the observed number of stars $N$ to determine what is required for a statistically significant detection. For our purposes, a detection at the level $n\sigma$ is defined as the mean lying $n\sigma$ away from zero. Figure \ref{fig:GPfit}(b) shows $\sigma$ as a function of $N$. Unfortunately, nearly $2\times10^4$ stars are required for a $3\sigma$ detection ($\sigma\approx0.5\times10^{-8}$~cm/s$^2$). This is a prohibitively large sample size given realistic quantities of observation time. However, we reiterate that $N$ represents the number of stars in an initial sample before target selection.

Next, we study the case where $v_{\textrm{noise}}$ contains only Gaussian white noise with $\sigma_{\textrm{WN}} = 10$~cm/s. This is a futuristic scenario we envision where data processing techniques developed in exoplanet astronomy have matured to the point of being able to effectively filter out the effects of correlated stellar noise.  At present, this is an unsolved problem, but it is being worked on very actively \cite{Dumusque2016,Dumusque2017,Fischer2016}.  Once stellar noise is removed, we reach the instrumental noise limit.

For this dataset of $N=10^3$ stars, we use a linear fit for each time series (i.e., fit only the acceleration component), and construct a histogram of the fitted slopes. To avoid broadening the histogram with Keplerian signals, we employ Lomb-Scargle periodograms \cite{Perryman2011} to identify Keplerian signals and reject time series containing planets and stellar companions (without a priori knowledge of their existence).

\begin{figure} 
\includegraphics[width=\columnwidth]{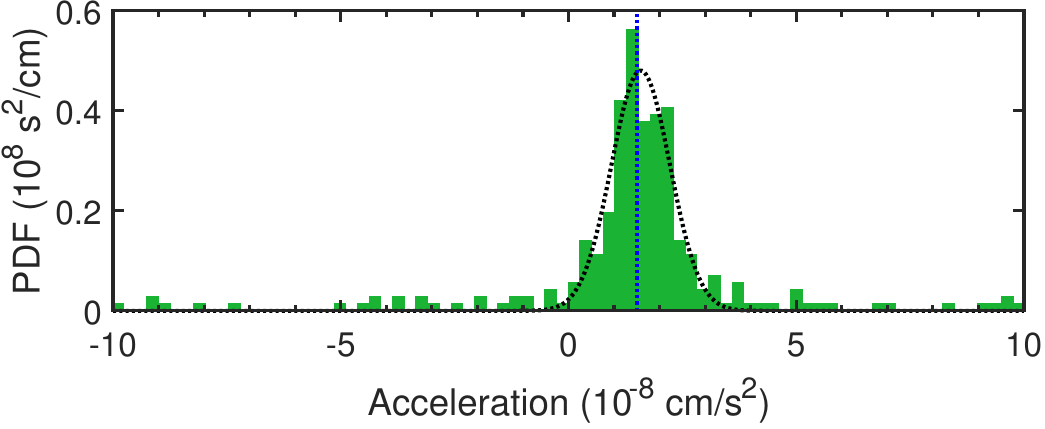} 
\caption{\label{fig:hist} Probability density function (PDF) of fitted stellar accelerations for $10^3$ stars, using a simple linear fit and filtering using periodogram power. The only source of noise included is 10 cm/s white noise. The dotted black curve is a Gaussian fit to the histogram (a guide to the eye). The injected acceleration signal of $1.5\times10^{-8}$ cm/s$^2$ is shown as a blue dotted line.}
\end{figure}

Setting a threshold on the maximum allowed periodogram power close to the maximum observed from a pure noise signal rejects significant Keplerian signals. Our threshold allows 277 stars to be examined, including all 132 lone stars (i.e., free of orbiting bodies) in the sample. Figure \ref{fig:hist} shows the result of such an analysis. Using this subset, we obtain a mean acceleration of 1.46(0.21)$\times10^{-8}$ cm/s$^2$. This result is consistent with the injected stellar acceleration signal of $1.5\times10^{-8}$ cm/s$^2$ and $\approx7\sigma$ away from a null result. Note that an acceleration measurement with 14\% uncertainty implies, using Eq.~\eqref{eq:rhoDM_unc}, a local DM density measurement with 38\% uncertainty.

We also test the dependence of the acceleration sensitivity with planet occurrence by doubling the mean number of planets per star, keeping all other parameters fixed. This also doubles, on average, the number of long-period (>10 year) planets, which are difficult to remove without sufficient coverage of the orbital period. With the added planets it takes roughly 4 times as many stars to reach the same detection significance.

Instead of rejecting companions via periodograms, they can be used to help fit Keplerian signals. This could enable reaching the same precision in the result with far less telescope time. In the future, we plan to simulate the target selection program in more detail, keeping track of the time overhead associated with following poor targets and also do a study of the stellar acceleration precision versus number of observations, as telescope time is an expensive resource.

\section{Conclusion}
In this Letter, we put forth the idea of using precision RV measurements to quantify the local DM density. More specifically, we propose to track the velocity of stars over time to extract their accelerations, thereby directly probing the local gravitational potential and foregoing the equilibrium assumption used with static measurements of stellar velocities. The exquisite RV precision and stability achievable with astro-comb wavelength calibrators and exoplanet spectrographs, combined with next-generation large telescopes, should make it feasible to measure stellar accelerations directly at the necessary level of $10^{-8}$ cm/s$^{2}$. Importantly, detecting accelerations at this level with an ensemble of 10$^3$ stars requires the reduction of the effect of stellar noise on RV measurements to <10 cm/s. In the future, measurements over a wide range of pointing directions could allow construction of a map of the gravitational potential of the Galaxy.

We conclude by emphasizing two key points. First, though the technical challenges are daunting for a large observing campaign to map stellar accelerations and constrain dark matter models, we believe such a program will become feasible in the next decade. Second, the dataset that would result from this effort, providing precision RVs from $\sim$~10$^3$ stars, would be rich and valuable for other areas of astrophysics: e.g., many long-period exoplanets and stellar companions would likely be detected and characterized in the process. It is our hope that the exciting prospects for exoplanet and stellar astronomy as well as dark matter physics will encourage others to consider the stellar acceleration idea and pursue it further.\\

\begin{acknowledgments}
We thank Nicholas Rodd and Katelin Schutz for useful discussions.  We also thank the Kavli Institute for Theoretical Physics, where the connection between the Harvard-Smithsonian and Michigan groups was first made.  The work of AR, NL, DFP, and RLW was supported in part by NASA award number NNX16AD42G.  The work of MB and BRS was supported in part by the DOE Early Career Grant DE-SC0019225.
\end{acknowledgments}

\bibliographystyle{apsrev4-1}

\end{document}